\PassOptionsToPackage{unicode}{hyperref}
\PassOptionsToPackage{hyphens}{url}
\PassOptionsToPackage{dvipsnames,svgnames,x11names}{xcolor}
\documentclass[
]{article}
\usepackage{amsmath,amssymb}
\usepackage{iftex}
\ifPDFTeX
  \usepackage[T1]{fontenc}
  \usepackage[utf8]{inputenc}
  \usepackage{textcomp} 
\else 
  \usepackage{unicode-math} 
  \defaultfontfeatures{Scale=MatchLowercase}
  \defaultfontfeatures[\rmfamily]{Ligatures=TeX,Scale=1}
\fi
\usepackage{lmodern}
\ifPDFTeX\else
\fi
\IfFileExists{upquote.sty}{\usepackage{upquote}}{}
\IfFileExists{microtype.sty}{
  \usepackage[]{microtype}
  \UseMicrotypeSet[protrusion]{basicmath} 
}{}
\makeatletter
\@ifundefined{KOMAClassName}{
  \IfFileExists{parskip.sty}{%
    \usepackage{parskip}
  }{
    \setlength{\parindent}{0pt}
    \setlength{\parskip}{6pt plus 2pt minus 1pt}}
}{
  \KOMAoptions{parskip=half}}
\makeatother
\usepackage{xcolor}
\usepackage[margin=1in]{geometry}
\usepackage{graphicx}
\makeatletter
\def\maxwidth{\ifdim\Gin@nat@width>\linewidth\linewidth\else\Gin@nat@width\fi}
\def\maxheight{\ifdim\Gin@nat@height>\textheight\textheight\else\Gin@nat@height\fi}
\makeatother
\setkeys{Gin}{width=\maxwidth,height=\maxheight,keepaspectratio}
\makeatletter
\def\fps@figure{htbp}
\makeatother
\usepackage{xcolor}
\usepackage{enumerate}
\usepackage{abstract}
\usepackage[utf8]{inputenc}
\usepackage[english]{babel}
\usepackage{latexsym}
\usepackage{graphicx}
\usepackage{amssymb}
\usepackage{amsmath}
\usepackage{amsthm}
\usepackage{dsfont}
\usepackage{booktabs}
\usepackage{cite}
\usepackage{setspace}
\usepackage[autostyle=true,german=quotes]{csquotes}
\usepackage[makeroom]{cancel}

\usepackage{bm}

\usepackage{marginnote}
\usepackage[title,toc,titletoc,page]{appendix}

\theoremstyle{plain}
\newtheorem*{prop*}{Proposition}
\newtheorem*{assump*}{Assumption}

\usepackage{float}
\restylefloat{table}

\definecolor{Red}{rgb}{0.6,0,0}
\definecolor{Green}{rgb}{0.1,0.5,0.1}
\ifLuaTeX
  \usepackage{selnolig}  
\fi
\usepackage[]{natbib}
\IfFileExists{bookmark.sty}{\usepackage{bookmark}}{\usepackage{hyperref}}
\IfFileExists{xurl.sty}{\usepackage{xurl}}{} 
\hypersetup{
  pdftitle={Regularized Adjusted Plus-Minus Models for Evaluating and Scouting Football (Soccer) Players using Possession Sequences},
  pdfauthor={Robert Bajons1; Kurt Hornik1},
  pdfkeywords={Rating systems, Sports analytics, Penalized
regression, Conic programming},
  colorlinks=true,
  linkcolor={red!50!black},
  filecolor={Maroon},
  citecolor={ProcessBlue!50!RoyalBlue},
  urlcolor={green!50!black},
  pdfcreator={LaTeX via pandoc}}

\title{Regularized Adjusted Plus-Minus Models for Evaluating and
Scouting Football (Soccer) Players using Possession Sequences}
\author{Robert Bajons\textsuperscript{1}\footnote{Corresponding Author.
  \href{mailto:robert.bajons@wu.ac.at}{\nolinkurl{robert.bajons@wu.ac.at}}} \and Kurt
Hornik\textsuperscript{1}}
\date{\scriptsize\textsuperscript{1} Vienna University of Business and
Economics, Institute for Statistics and Mathematics}

\begin{document}
\maketitle
\begin{abstract}
This paper presents a novel framework for evaluating players in
association football (soccer). Our method uses possession sequences,
i.e.\textasciitilde sequences of consecutive on-ball actions, for
deriving estimates for player strengths. On the surface, the methodology
is similar to classical adjusted plus-minus rating models using mainly
regularized regression techniques. However, by analyzing possessions,
our framework is able to distinguish on-ball and off-ball contributions
of players to the game. From a methodological viewpoint, the framework
explores four different penalization schemes, which exploit
football-specific structures such as the grouping of players into
position groups as well as into common strength groups. These four
models lead to four ways to rate players by considering the respective
estimate of each model corresponding to the player. The ratings are used
to analyze the 2017/18 season of the Spanish La Liga. We compare
similarities as well as particular use cases of each of the penalized
models and provide guidance for practitioners when using the individual
model specifications. Finally, we conclude our analysis by providing a
domain-specific statistical evaluation framework, which highlights the
potential of the penalized regression approaches for evaluating players.
\end{abstract}

\section{Introduction}\label{introduction}

Evaluating player performances in professional football, sometimes
referred to as soccer, has attracted a lot of interest in the last
decade. Due to the low-scoring and complex nature of the sport,
accurately measuring a player's contribution to the game is a
challenging task. However, evaluating individual performances in team
sports is a fundamental task of increasing importance. While there is an
external interest from fans \citep{McHale12}, media \citep{Kharrat20},
and the betting industry \citep{OeML23} another particularly important
aspect nowadays is to use player ratings to inform scouting and
retention decisions of teams while being compliant with budget
constraints \citep{Pap19,PH21}.

\par

In football, the decision-making process is traditionally realized by a
qualitative assessment of players by professionals such as coaches,
managers, or scouts. However, lately, there have been many attempts and
advances toward a more objective approach to quantifying the impact of
players on the game. For example, traditional statistics such as goals
or pass completion percentage, which have been used extensively to
analyze the strength of players, have been replaced by metrics such as
expected goals (xG, cf.\textsubscript{e.g.}\citealp{RD20,AB21}) or
passing models based on statistical approaches
(\citealp{SM16,Hvattum_19,Haland20}). In general, these novel
data-driven methods can be categorized into two distinct groups:
bottom-up and top-down rating systems
(cf.\textasciitilde{}\citealp{HvattumGelade21}). The former assesses
player performances by assigning values to each action performed and
then aggregating them for each player over the course of a relevant
period (i.e.\textasciitilde a match or a season). Top-down ratings on
the other hand evaluate players by breaking down the whole team's
performance and distributing credit onto the players involved.
Typically, bottom-up approaches use very granular data on an event level
in order to assign values to either specific actions such as shots (xG)
or passes or more generally any type of action
(cf.\textasciitilde{}\citealp{DBHD19}). While bottom-up ratings are
therefore able to assess a player's quality down to each particular
action performed by the athlete, they however lack the ability to
account for team strengths as well as off-ball actions by teammates. In
contrast, top-down ratings such as plus-minus (PM) models (see
\citealp{HvattumPM19} for an overview) rely only on information gained
from specific segments of a match, which in general is much less
detailed than event stream data. These approaches are by design not able
to account for specific actions taken by a player. They are, however,
able to incorporate effects of teammates.

\par

In this work, we present a novel framework for evaluating player
strength. On the surface, the methodology is similar to regularized
adjusted plus-minus (PM) models as we mainly use penalized regression
models in order to infer a rating of players. However, we extend the
classical PM approach in two ways. First, we separate matches into
possessions, which are sequences of consecutive on-ball actions. Thus,
we consider much more granular segments of the game than existing PM
models, where a segment is defined as one where the same set of players
is on the field (\citealp{Kharrat20,HvattumPM19}). The segmentation in
possessions leads to a data structure where it is possible to separate
effects of being part of a possession, i.e.\textasciitilde actually
performing on-ball actions (direct involvement), from effects of only
being on the pitch (indirect involvement). From a domain-specific
viewpoint, this is highly attractive as it is well-known that off-ball
actions have a substantial impact in soccer (cf.~\citealp{S18,BFC21}).
Using possessions data, it is also necessary to adjust the dependent
variable that measures the outcome of a segment. In the classical
regularized adjusted PM approach the (normalized) goal difference in a
segment is used. Since possessions are way shorter segments, goal
difference is not appropriate as response variable. Instead, we focus on
whether the possession ends in a goal or not as our outcome variable.
Second, we provide a variety of penalization structures, which have not
been considered in this context. Most of the classical plus-minus models
use a ridge regression penalty, i.e.\textasciitilde an \(L_2\)-type of
penalty on the regression coefficient
(cf.\textasciitilde e.g.~\citealp{Kharrat20}). In a more innovative
approach, \citet{PH21} adjust the penalty term by accounting for an age
effect as well as a similarity effect within the shrinkage procedure.
However, to the best of our knowledge, there has been no comparison or
analysis of penalization structures beyond (adaptions of) the ridge
penalty. We analyze several regularization structures that exploit the
implicit grouping of players that arises from the different position
groups on the field. In this way, our player evaluations are able to
provide interesting and novel insights into player strengths where
different penalties may be used for different use cases from
practitioners. Finally, we conduct a statistical evaluation of our
methods specifically suited for football based on validity experiments
from \citet{HvattumGelade21}. Our methods are compared to reference
models using different kinds of data (ELO team ratings) and another
modeling approach using a debiased machine learning approach
(\citealp{Bajons23}).

\par

The remainder of the paper is structured as follows. In Section
\ref{sec-data} we describe the data used for this research and the
necessary preprocessing steps. In Section \ref{sec-meth}, we provide
details on the novel regularization structures and models used to
determine player strength. Statistical descriptions are accompanied by
managerial implications of each model. Section \ref{sec-res} presents a
comparison of the ratings as obtained from the different models and our
approach to statistically evaluate the methods. Finally, we conclude the
paper with a discussion in Section \ref{sec-concl}.

\section{Data}\label{sec-data}

We use raw event stream data from the Spanish La Liga season of 2017/18,
openly available via
\href{https://figshare.com/collections/Soccer_match_event_dataset/4415000/5}{Figshare}
(\citealp{Event19}), as a basis in order to create a dataset of
possessions. A possession is defined as a sequence of consecutive
on-ball events, which ends either with the opponent team gaining the
possession or by an action of the referee, i.e.\textasciitilde a foul,
the ball leaving the pitch, or the end of a period.

\par

As the main goal is the valuation of players, it is necessary to extract
information on which player was involved in a possession sequence.
Extracting offensive player involvement in possessions is
straightforward from the data, however, information on the opposing
players is desirable as well. Since event stream data only provides
details on the on-ball actions, it is difficult to extract meaningful
opponent player information. One way to incorporate such knowledge is to
record all players on the field during a possession. Thus, we consider a
regression matrix \(X_{mod} = (X_{N_{inv}},X_{N_{of}})\) with
\(N = N_{inv}+N_{of}\) columns and \(N_p\) rows. Each of the involvement
columns \(N_{inv}\) is a binary variable for a specific player, that is
1 whenever the player was involved in the possession and 0 otherwise.
Each of the \(N_{of}\) columns indicates whether a specific player was
on the field for the possession team (indicated by 1) or on the field
for the defensive team (indicated by \(-1\)). In this work,
\(N_{inv} = N_{of} = 556\) corresponds to the number of players in the
data set. Note that because we are interested in evaluating players we
do not extract possession-specific variables. The data includes spatial
and temporal features such as the length and speed of possessions, which
may help predict the success of these. However, the underlying
assumption is that spatial or temporal effects are due to the players
being involved in the possession as well as on the field. That is, while
it may be the case that quick but long possessions increase the chance
of a successful outcome it is due to the quality of players involved and
on the field that it is even possible to execute such an attack.

\par

The number of rows \(N_p\) depends on the number of possessions
analyzed. As we are interested in deriving a player rating it makes
sense to only consider valuable possessions, where we define a
possession as valuable if it ends in the last third of the pitch
(i.e.\textasciitilde sufficiently close to the opponent's goal).
Following our domain-specific intuition, we further only consider
possessions with at least 3 consecutive actions. The only exceptions to
this rule are possessions that start with a free kick or a corner kick,
as they are relevant for scoring goals and often comprise only two
actions (e.g.\textasciitilde the corner kick and a header). This, in
particular, eliminates all the penalty kicks in our data. It is common
practice to exclude these highly specific situations
(\citealp{RD20,AB21}). In most other circumstances, short possessions
are a result of duels for the ball and are thus not necessarily directly
relevant for detecting a player's influence on scoring a goal from a
possession. In total, the data contains \(N_p = 53233\) possessions.

\par

Finally, a relevant response variable is extracted, which allows us to
analyze the effect of players on possessions. Since goals are the most
important part of the game of football, a goal indicator is taken as
outcome variable. As football is a low-scoring game, one drawback is
that this outcome variable is quite imbalanced. In our data, only 1.3 \%
of the possessions end in a goal.

\par

On average the possessions contain \(\approx 7.5\) consecutive actions
with a median of \(6\) consecutive actions. The average number of direct
involvements for players in possession is 394.8. The distribution of the
number of involvements is skewed to the right, which intuitively
reflects the idea that only a few outstanding players have a lot of
involvements whereas especially defenders are not as often involved in
possessions. It is important to mention, that from a domain-specific
viewpoint, several possible adjustments to the data setup could be
considered. While we discuss some of these in more detail in Section
\ref{sec-concl}, we emphasize that for the purpose of presenting the
benefits of our methods, the setup is well-suited.

\section{Methods and practical implications}\label{sec-meth}

With the possessions data at hand, the most naive and straightforward
approach to deriving a player rating would be to use classic regression
techniques and fit a generalized linear model to the data. Then the
models' regression coefficients could be used as a strength estimate of
each player. However, as many studies point out
(\citealp{Taddy13,SH15,HvattumPM19,Kharrat20}), such an approach is not
expected to work in the high-dimensional and sparse setup at hand, due
to classical issues such as multicollinearity, overfitting and the high
imbalance in outcome and independent variables.

\par

In the following, we discuss several approaches to deriving a rating of
football players from the processed event data that overcome the
previously mentioned problems. The main class of models is regularized
or penalized regression, often referred to as adjusted plus-minus
models, which have been studied in the context of sports player ratings
by many authors previously (e.g.,
\citealp{Taddy13,HvattumPM19,Kharrat20}). We adapt the methodology to
our data setup and discuss various forms of penalized regression models.
For each of the models, first, a statistical description is provided and
secondly, a detailed domain-specific interpretation of the penalization
structure as well as implications and use cases for practitioners are
discussed.

\hypertarget{subsec-ridge}{%
\subsection{Ridge regression}\label{subsec-ridge}}

One of the earliest approaches for regularization dating back to
\citet{ridge70} is ridge regression (sometimes also referred to as
Tikhonov regularization). The ridge estimator is given as
\begin{equation}
\label{eq-ridge_est}
\hat{\beta}_{ridge} = \underset{\beta}{\operatorname{argmin}} \text{ } \ell(Y;X,\beta) + \lambda ||\beta||^2_2,
\end{equation} where \(\ell(Y;X,\beta)\) represents the (negative)
log-likelihood function for the binomial response variable \(Y\) in the
usual generalized linear model framework that we work in. \% In the case
of a Gaussian likelihood function, an explicit solution for the above
problem can be derived (see e.g.~\citealp{ESL13}). For generalized
linear models, the problem is typically formulated as a weighted
least-squares problem which is then solved in an iterative fashion
(cf.~\citealp{glmnet10}). The solution \(\hat{\beta}_{ridge}\) of
problem \eqref{eq-ridge_est} is quite appealing as it shrinks the
coefficients towards each other, where more shrinkage is put on the
columns that explain less variance in the regression matrix \(X\). That
is, as we expect the response variable, i.e.\textasciitilde goals scored
from possession, to vary most in the directions of important players,
this approach shrinks regular players more than strong (or weak)
players. However, players with very similar strength will likely be
shrunken towards a similar value. This is quite desirable from a
perspective of a rating, as every player gets a rating that reflects his
strengths in relation to every other player on the field.

\par

Another particularly nice feature arises from the data setup. As the
regression matrix \(X = (X_{N_{inv}},X_{N_{of}})\) consists of direct
involvement columns \(N_{inv}\) as well as on-field indicator columns
\(N_{of}\), there are two sources of information for each player \(i\)
represented by two estimates \(\hat \beta_{i,inv}\) and
\(\hat \beta_{i,of}\). \%The (shrunken) estimate \(\hat \beta_{i,inv}\)
of player \(i\) represents a direct involvement strenght, while
\(\hat \beta_{i,of}\) characterized the effect of player \(i\) being on
the field. \%One is the direct information of how much a player \(i\)
contributes to specific possessions when involved, given by the
corresponding (shrunken) estimate representing player \(i\)'s column in
the sub-matrix \(X_{N_{inv}}\). The other one is the indirect
contribution to possessions, given by player \(i\)'s estimate from the
sub-matrix \(X_{N_{of}}\). From a practical perspective this is nicely
interpretable: A player, even though potentially not involved in the
possession, usually nevertheless has an influence on the game. A
threatening forward for example may create space for other players,
simply because the opposing team tries to cover the player. On the other
hand, a strong goalkeeper who deflects a lot of shots also clearly
influences offensive possessions of the opposing team. Such effects may
be measured from the estimate of indirect involvement
\(\hat \beta_{i,of}\). The total strength of a player \(i\) could then
e.g.\textasciitilde be measured by the sum of both contributions. In
Section \ref{sec-res} we discuss several methods of assembling the two
strength estimates of a player in order to obtain an overall strength
estimate of players.

\hypertarget{subsec-grouplasso}{%
\subsection{Group lasso and exclusive lasso}\label{subsec-grouplasso}}

Another common form of regularization is lasso regression, where instead
of imposing an \(L_2\)-norm constraint on the coefficients \(\beta\) (as
is done in ridge regression), an \(L_1\)-norm penalty is used. The
estimate is thus defined as \begin{equation}
\label{eq-lasso_est}
\hat{\beta}_{lasso} = \underset{\beta}{\operatorname{argmin}} \text{ } \ell(Y;X,\beta) + \lambda ||\beta||_1.
\end{equation} This penalty allows for implicit variable selection, as
some of the estimates are not only shrunken towards zero but set exactly
to zero (cf.\textasciitilde e.g.~\citealp{ESL13}). While such an effect
is desirable in many situations, as variables with less influence are
omitted from the regression model and only influential variables retain
a value different from zero, in our setup it is suboptimal. Even though
it may be possible to detect strong (or weak) players from the resulting
coefficients, the majority of them will be set to zero and no
interpretation is possible anymore. This is especially inconvenient for
the case of recruiting players, as a team is usually looking for a
tradeoff between a strong player and an affordable player. Furthermore,
the appealing effect of interpreting direct and indirect involvement to
possession as in the ridge regression case is not as easy to employ. In
the following, we discuss variants of the lasso penalty, which are more
useful for our player evaluation framework.

\par

A particularly popular variant is the group lasso penalty (see
e.g.~\citealp{SGL13}) given as \begin{equation}
\label{eq-group_lasso_est}
\quad \min_\beta \ell(Y;X,\beta)+\lambda \sum_{l=1}^m \sqrt{p_l}\left\|\beta^{(l)}\right\|_2,
\end{equation} where the predictor variables are divided into \(m\)
groups. In equation \eqref{eq-group_lasso_est}, \(\ell(Y;X,\beta)\)
represents a scaled version of the negative log-likelihood and \(p_l\)
are weights representing the length of the \(l\)-th group specific
estimate \(\beta^{(l)}\) (cf.~\citealp{SGL13}). This penalty results in
a shrinkage of the coefficients of complete groups to zero,
i.e.\textasciitilde on a group level it performs shrinkage similar to
lasso. However, within each group, it does not perform variable
selection, leading to a similar effect as ridge regression. Thus, if
each group is of size one the group lasso results in the classical
lasso. On the other hand, if there is only one group the penalty is
similar to ridge regression.

\par

In our application of rating football players, there is a natural
grouping of players as usually, each player plays a specific position.
While these positions may vary slightly depending on specific matches
and tactics used, players can be categorized into at least 4 broad
categories: Forward, midfielder, defender, and goalkeeper. The group
lasso enables us to distinguish between players in each of these groups.
On a group level, whole groups are shrunken simultaneously. If there
were one irrelevant group (say for example the goalkeepers, whose main
task is not to be involved in possession of high threat), then this
group would be shrunken towards zero. Within a group however, the
ridge-like penalty shrinks players of similar strength to a similar
value, while less shrinkage is applied to strong (or weak) players. From
an interpretational viewpoint, this is highly desirable. As the tasks on
the field of each individual group vary, we do not expect comparisons
between two groups, e.g.\textasciitilde forwards and defenders, to be
fair and instead might be more interested in within-group comparisons.
Another dimension arises again from the distinction between direct and
indirect involvement. Similar to the ridge regression case one can
combine both aspects for the groups and perform within-group analyses.

\par

Another variation of the lasso is the exclusive lasso (see
e.g.~\citealp{ExL17}), which solves the problem \begin{equation}
\label{eq-excl_lasso_est}
\quad \min_\beta \ell(Y;X,\beta)+\frac{\lambda}{2} \sum_{l=1}^m \left\|\beta^{(l)}\right\|_1^2,
\end{equation} where \(\ell(Y;X,\beta)\) again represents a scaled
version of an exponential family likelihood function and the regressors
can be separated into \(m\) different groups. The behavior of this
penalty is analyzed in detail in \citet{ExL17}. As opposed to the group
lasso, it can be obtained from equation \eqref{eq-excl_lasso_est} that
between groups an \(L_2\) (i.e.\textasciitilde ridge) penalty is
applied, while within each group an \(L_1\) (i.e.\textasciitilde lasso)
penalty is applied. That is, on a group level the penalty does not imply
shrinkage of complete groups, i.e.\textasciitilde in each group at least
one variable will be non-zero. Within groups however only the most
relevant variables will be selected, i.e.\textasciitilde the penalty
tries to set most of the group members to zero as is usual for the lasso
penalty. In the extreme case, exactly one non-zero variable of each
group will be selected.

\par

The application-specific use case of the penalty reveals more potential
in scouting rather than in rating players. While one could again
categorize the players into the above 4 categories based on their broad
position on the pitch and analyze which players are particularly strong
(resp. weak) in their respective group, the exclusive lasso motivates
more interesting selection procedures. A particular use case would be a
team that is interested in signing players who play a specific position
and who are at a certain age level or below a certain market value. In
this case, an individual group could be formed and the exclusive lasso
would result in the selection of a few strong (resp. weak) players that
fall into the prespecified type, while at the same time being able to
account for the usual groups as defined by e.g.\textasciitilde player
position. More details on using the exclusive lasso for scouting are
found in the supplementary materials for this work.

\hypertarget{subsec-genlasso}{%
\subsection{Generalized lasso}\label{subsec-genlasso}}

In the previous section, we discussed penalizing structures on
prespecified groups and their usage in analyzing football players.
However, a different approach is to set up the penalizing structure such
that implicit grouping is achieved. In the context of analyzing sports
teams, \citet{RL12} proposed the ranking lasso, given by the problem
\begin{align}
\label{eq-rank_lasso_est1}
\min_{\beta} \ell(Y;X,\beta)+\lambda \sum_{i<j}^N w_{ij}|\beta_i-\beta_j|.
\end{align} The penalty term here is a variation of the lasso that
allows for the grouping of the estimated abilities of players into
several equivalence classes. More generally the problem can be written
as \begin{align}
\label{eq-gen_lasso_est1}
\min_{\beta} \ell(Y;X,\beta)+\lambda ||D\beta||_1,
\end{align} where \(D\) is a matrix that allows for individual
specification of lasso-type penalties. The penalization structure in
equation \eqref{eq-gen_lasso_est1} is termed generalized lasso
(cf.\textasciitilde e.g.~\citealp{GenLasso11}) and includes the ranking
lasso as well as the more well-known fused lasso (\citealp{FL05}).

\par

Applying the generalized lasso penalty in the player evaluation setting
is not completely straightforward. The aim is to cluster similar player
contributions together, however as the structure of the regression
matrix is \(X = (X_{N_{inv}},X_{N_{of}})\), it does not make sense to
consider the simple ranking lasso penalty. Instead, a more natural
approach is to consider direct involvements separated from indirect
ones. We first examine the structure of the matrix \(D_{N_{inv}}\) for
the coefficients corresponding to one of the sub-matrices of \(X\),
namely \(X_{N_{inv}}\). The goal is to employ a penalty such that
similar strength estimates are shrunken together, i.e.\textasciitilde to
apply the ranking lasso to the estimates of these columns. Thus the
matrix \(D_{N_{inv}}\) can be expressed as a block matrix \begin{align}
D_{N_{inv}} = (A,B,C) = \left(\begin{array}{ccc}
A_1 & B_1 & C_1 \\
\vdots & \vdots & \vdots \\
A_{N-1} & B_{N-1} & C_{N-1}
\end{array}\right).
\end{align} containing three blocks
\(A^{\top} = (A_1^{\top},\dots,A_{N-1}^{\top})\),
\(B^{\top} = (B_1^{\top},\dots,B_{N-1}^{\top})\) and
\(C^{\top} = (C_1^{\top},\dots,C_{N-1}^{\top})\). Each \(A_i\) is a
matrix of zeroes of dimension \((N_{inv}-i) \times (i-1)\), each \(B_i\)
is a matrix (actually a vector) of ones of dimension
\((N_{inv}-i) \times 1\) and each \(C_i\) the negative identity matrix
of dimension \(N_{inv}-i\). The same penalization structure can be
applied to the coefficients of indirect involvement,
i.e.\textasciitilde these are also shrunken according to the ranking
lasso penalty and thus \(D_{N_{inv}} = D_{N_{of}}\). In order to obtain
the full penalization matrix \(D\), the two sub-matrices \(D_{N_{inv}}\)
and \(D_{N_{of}}\) need to be concatenated suitably: \begin{align}
D = \left(\begin{array}{cc}
D_{N_{inv}} & \boldsymbol{0}  \\
\boldsymbol{0} & D_{N_{of}}
\end{array}\right).
\end{align} Indeed one can think of different approaches for the
penalization matrix \(D\), for example, if \(D_{N_{of}}\) is the
identity matrix, this would correspond to a classical lasso penalty on
the indirect involvement coefficients. For the application to player
evaluation, it makes however sense to stick to a structure where the
coefficients of direct involvements in possessions are separated from
the ones of indirect. In this way, this penalization approach allows for
the selection of clusters of players with equal strength for on-ball and
off-ball involvements separately. This may be of interest, when the goal
is to identify players of similar strength and playing style compared to
a reference player.

\section{Computational details and results}\label{sec-res}

\subsection{Implementation}\label{implementation}

All of the models discussed in Section \ref{sec-meth} are applied to the
event data set described in Section \ref{sec-data}. All computations are
performed using the \texttt{R} statistical programming language
(\citealp{R22}) and we make use of several contributed packages. For the
ridge regression, the \texttt{glmnet} function from the correspondent
package (\citet{glmnet10} and \citealp{glmnet2}) is used, the group
lasso is fitted using the package \texttt{SGL} (\citealp{SGL_R19}) and
for the exclusive lasso the \texttt{ExclusiveLasso} package
(\citealp{ELR18}) is provided. In all cases, we use a suitable
cross-validation procedure in order to select the optimal shrinkage
parameter \(\lambda\). For the generalized lasso, there is no \texttt{R}
package readily available. Although there exists the \texttt{genlasso}
package (\citealp{GenLasso22}), this only provides an implementation for
the case of Gaussian outcome variables.

\par

We briefly describe our implementation for the binomial logistic
regression. We use modern conic programming principles to solve the
problem. In particular, equation \eqref{eq-gen_lasso_est1} can be
written as \begin{align}
\label{eq-ada_gll}
\min_{\boldsymbol{\beta}}  -\big(\sum_{i = 1}^{n} y_i \log(h_{\boldsymbol{\beta}}(x_i))+(1-y_i)\log(1-h_{\boldsymbol{\beta}}(x_i))\big)+\lambda \sum_{j = 1}^m w_{j}|D_j|,
\end{align} with logistic function
\(h_{\beta}(x) = 1/\big(1+\exp(-\beta^{\top} x)\big)\) and pair-specific
weights \(w_{ij}\). In this case, \(D_j\) is the \(j\)-th component of
\(D\boldsymbol{\beta}\), with \(D\in \mathbb{R}^{m \times p}\). This
problem can be reformulated as a conic program by expressing it in its
equivalent epigraph form (see \citealp{BV04}). Specifically,
\eqref{eq-ada_gll} can be written as \begin{equation}
\label{CCOP}
\begin{aligned}
 \min_{(\boldsymbol{r},\boldsymbol{\beta},\boldsymbol{s},\boldsymbol{t},\boldsymbol{z}_1,\boldsymbol{z}_2)} & \sum_{i = 1}^n t_i +\lambda r \\
 \text{s.t.} & \left(z_{i1}, 1, u_i-t_i\right) \in K_{\exp}, \\
& \left(z_{i2}, 1,-t_i\right) \in K_{\exp }, \\
& z_{i1}+z_{i2} \leq 1, & i = 1,\dots,n, \\
& -s_j \le D_j \le s_j, & j = 1,\dots,m, \\
& r - w_1D_1 - \dots - w_mD_m \ge 0.
\end{aligned}
\end{equation} In the above formulation, the variables
\(\boldsymbol{r},\boldsymbol{\beta},\boldsymbol{s},\boldsymbol{t},\boldsymbol{z}_1,\boldsymbol{z}_2\),
and \(\boldsymbol{u}\) are auxiliary variables. The initial problem is
thus rewritten as a conic programming problem on an augmented set of
variables
\(\Theta = (\boldsymbol{r},\boldsymbol{\beta},\boldsymbol{s},\boldsymbol{t},\boldsymbol{z}_1,\boldsymbol{z}_2)\).
Such a problem can be solved using a convex optimization solver of
choice. Note that in order to solve \eqref{CCOP}, only two types of
conic cones are necessary, namely the linear cone and the exponential
cone. We use the \texttt{R}-package \texttt{ROI} (\citealp{ROI20}) which
allows to nicely handle such types of optimization problems in a
flexible manner.

\subsection{Comparing ratings}\label{subsec-comp}

We provide a comparison of the penalization methods described in section
\ref{sec-meth}. The aim here is to highlight idiosyncrasies as well as
commonalities of each approach. A more thorough analysis containing
specific rankings, player analyses, and further results can be found in
the supplementary material of this work.

\begin{figure}
  \centering
  \includegraphics[width=1\textwidth]{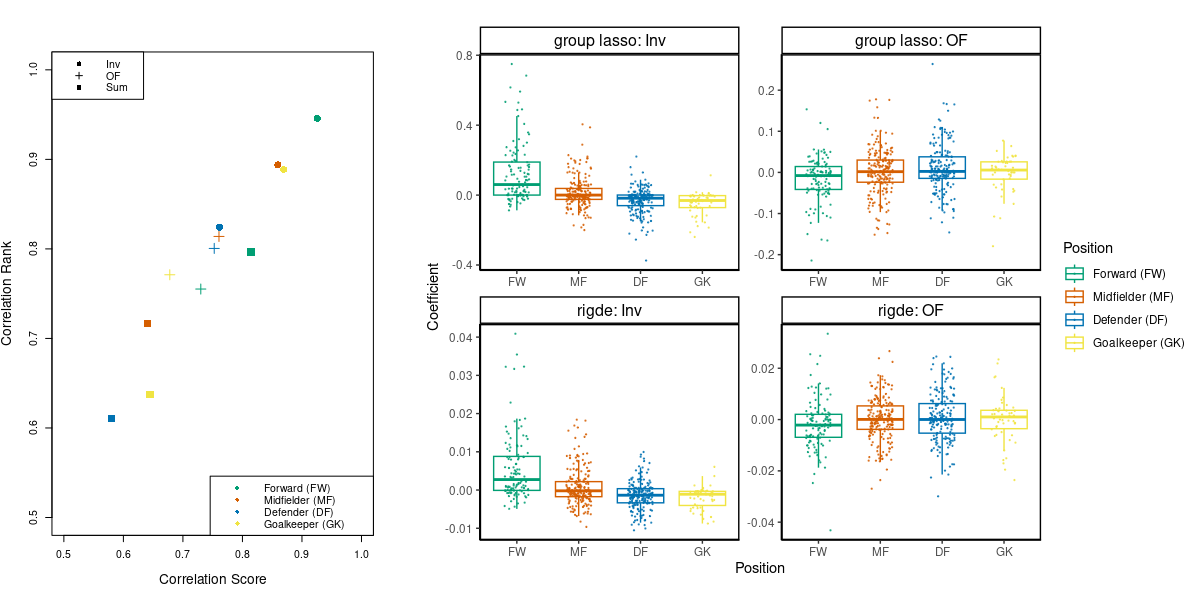}
  \caption{Comparison of the results from the ridge penalty and the group lasso penalty. Left: Correlation between the score of each group on the x-axis vs. correlation between the rank of each group on the y-axis. Groups are categorized by type of involvement (Inv, OF, sum) and position (FW, MF, DF, GK) Right: Boxplots of the estimates of each player in the 4 position groups under the group lasso and ridge penalty, separated by type of involvement.}
  \label{fig-ridge_vs_gl}
\end{figure}

First, we compare the ratings obtained from ridge regression to the ones
obtained from the group lasso. For the latter, the columns are split
into eight groups. The players' direct and indirect involvements are
grouped into four player positions (forward, midfielder, defender,
goalkeeper), resulting in four direct involvement groups and four
indirect involvement groups. In general, the effect of the group lasso
penalty is similar to the ridge regression as is visually confirmed by
Figure \ref{fig-ridge_vs_gl}. Offensive players (forwards and
midfielders) load higher on the coefficient of direct involvement (Inv)
and defensive players (defenders and goalkeepers) tend to load higher on
the coefficient of indirect (on-field) involvement (OF) as is shown in
the boxplots of Figure \ref{fig-ridge_vs_gl}. However, it seems that the
effects are more pronounced, especially for the direct involvements.
This is reasonable since the idea of the group lasso is to shrink whole
groups and thus less relevant groups, such as the defensive players, are
penalized more severely. The left part of Figure \ref{fig-ridge_vs_gl}
analyses the relationship between the ridge estimates and the group
lasso estimates in more detail. In terms of correlation between the two
rating methods, it seems that there is the most agreement for the
coefficients of direct involvements rather than the indirect
involvements. Furthermore, the ratings of the offensive players are more
strongly correlated than the ones for the defensive players. Overall,
the coefficients seem to be similar in both types of involvement in
possessions with correlations of 0.7 or greater for all position groups.
When trying to obtain a universal score for players, an intuitive
approach would be to consider the sum of the coefficients from direct
and indirect involvements. However, as seen in Figure
\ref{fig-ridge_vs_gl} (left panel), the correlation between the two
ratings is the lowest for the sum, suggesting that it may be worth
studying different approaches of combining the two individual
contributions to possession rather than simply summing them. Finally,
interesting analyses can be done when comparing the Inv and OF for
specific players and position groups, which we discuss in more detail in
the supplement.

\par

\begin{figure}
  \centering
  \includegraphics[width=1\textwidth]{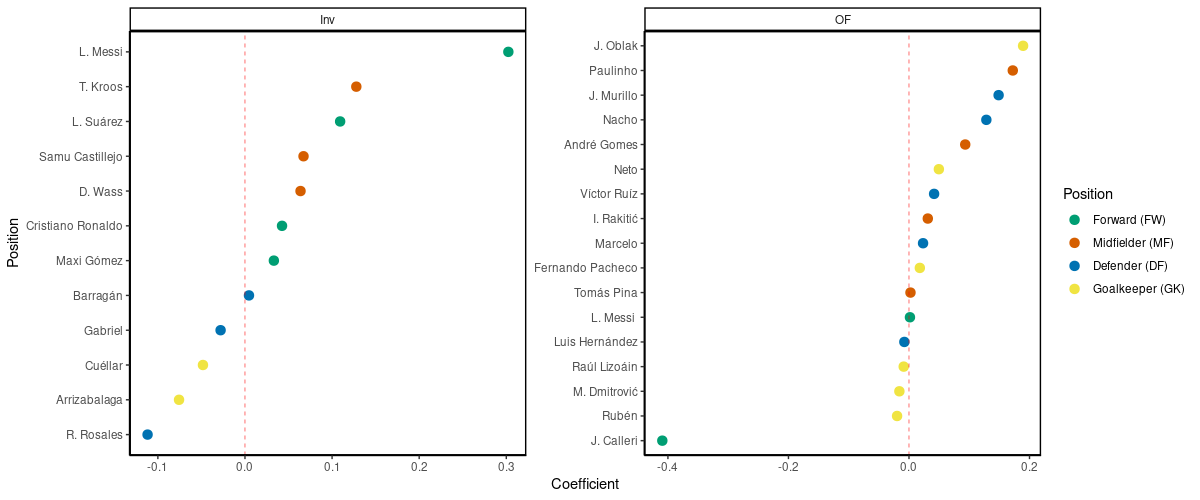}
  \caption{Estimates of the players selected from the exclusive lasso penalty. Left: Players selected from the 4 direct involvement (Inv) position groups. Right: Players selected from the 4 indirect involvement (OF) position groups.}
  \label{fig-excl_lasso}
\end{figure}

Next, we discuss the results from the exclusive lasso. To allow for
comparability, we use the same grouping as for the group lasso. Recall,
that the exclusive lasso performs lasso-like shrinkage within each group
but selects at least one relevant representative from each group. Figure
\ref{fig-excl_lasso} displays all selected players from the four direct
involvement (Inv) as well as the four indirect involvement (OF) groups.
The general pattern is similar to the two previous approaches. The
offensive groups load stronger on Inv, with more players selected and
higher coefficients (left panel of Figure \ref{fig-excl_lasso}). For OF,
defensive players and groups score higher (right panel of Figure
\ref{fig-excl_lasso}). As mentioned in Section \ref{subsec-grouplasso},
the exclusive lasso has its advantage in scouting, which would require a
different setup of the groups. We refer the interested reader to the
supplementary materials, where we present specific applications and
discuss benefits and pitfalls when using this penalization variant.

\begin{figure}
  \centering
  \includegraphics[width=0.8\textwidth]{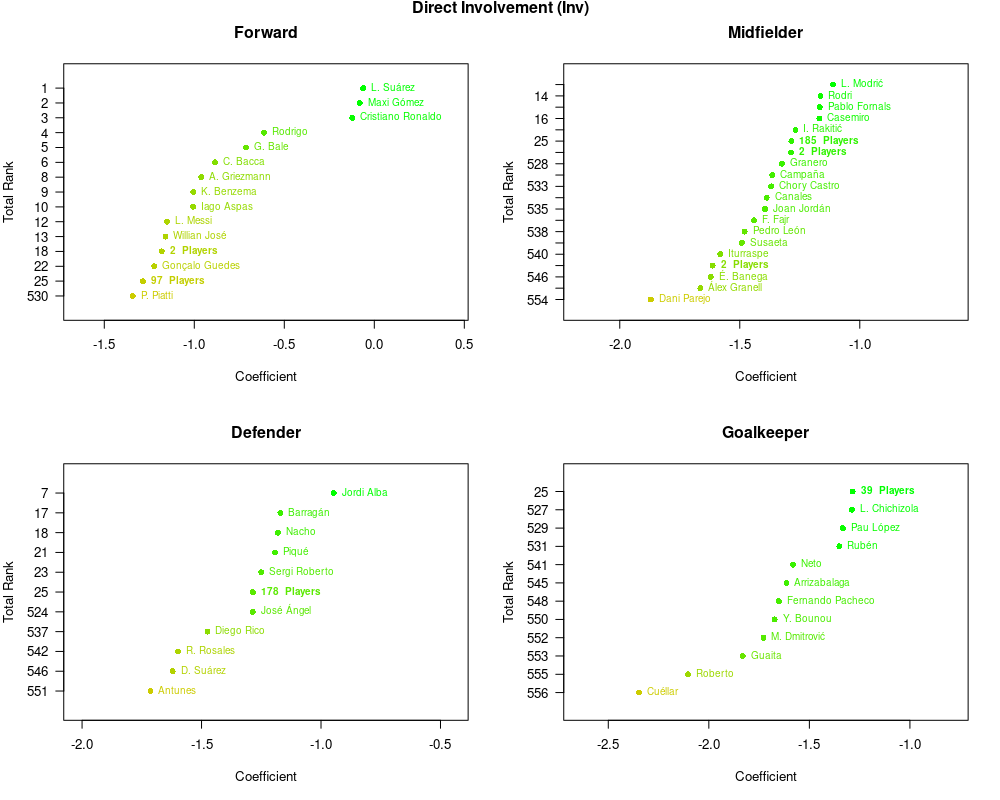}
  \caption{Results for each of the 4 position groups with regards to direct involvement (Inv) under the generalized lasso penalty. Coefficient groups are represented by their coefficient estimate (x-axis) and their total rank (y-axis). Groups containing only one player are marked by the name of the players.}
  \label{fig-gen_lasso_inv}
\end{figure}

Finally, we analyze the results from the generalized lasso. In contrast
to the group lasso and the exclusive lasso, this penalty allows for
implicit grouping of players. Figure \ref{fig-gen_lasso_inv} shows the
results for direct involvement coefficients. For comparability, we
separate the players again into the usual four position groups. The main
message is again consistent with the previous approaches: Offensive
players perform better with regard to direct involvement than defensive
players. We omit the results for indirect involvement and refer again to
the supplementary material. Interestingly, the grouping formed is rather
dense, i.e.~outstanding players (positively and negatively) form their
own groups while average players are bundled in one large group,
suggesting that the distinction between these players is difficult even
though theoretically possible within the ridge or group lasso framework.

\subsection{Validity of ratings}\label{subsec-validity}

Having derived a statistically motivated rating, an equally important
task is to objectively analyze its validity. One criterion for a
reasonable evaluation system of player strength discussed by
\citet{Franks16} and \citet{HvattumGelade21} can be summarized under the
term ``relevance'\,'. A metric is relevant, if there is a relationship
between the metric and an outcome of interest. In terms of football, the
most interesting outcome is the result of a game. Thus, a relevant
metric can establish a predictive relationship between a player's
strengths and the result of a match.

\par

The validity checks conducted in this paper follow the conceptual idea
of \citet{HvattumGelade21}. \%As we only work with data from the La Liga
data from 2017/18, the framework is adjusted similarly to the one in
\citet{Bajons23}. In particular, match outcome data of the 2017/18
season is divided into a training set containing the first 280 matches
and a test set containing the remaining 100. For each, a strength
estimate of the competing teams is derived as the average over the
strength values of players playing for the team. As the penalized
regression models contain two different measures of player strength (Inv
and OF), two distinct methods for concatenating them are applied. First,
the sum of both effects is used as discussed in Section
\ref{subsec-comp}. Second, the estimates are standardized such that they
range from 0 (indicating the weakest player) to 1 (indicating the
strongest player) for the two contributions. Then, we average them to
determine a player's strength. Furthermore, we adapt the grouping for
the exclusive lasso. As the exclusive lasso by design attempts to set
most of the players in a particular group to zero, considering only the
four position groups, as is done in
Section\textasciitilde{}\ref{subsec-comp}, results in impractically
strong sparsity. A more representative grouping for predicting match
results considers the four position groups for all teams in our data. In
this way, the algorithm selects the strongest (and/or possibly weakest)
players on each position group for every team.

\par

Two state-of-the-art models are considered for the prediction of match
results. First, a bivariate Poisson (BP) model as discussed by
\citet{KN03}, which models the number of goals by the home and away team
jointly based on covariates. Second, an ordered logistic regression
(OLR) model as proposed in \citet{ArntzenHvattum21}, which uses an
ordered categorical variable with three levels as response, namely home
win (\(H\)), draw (\(D\)) and away win. In particular, for both
prediction models, only one covariate is used to analyze the respective
outcomes, namely the difference in strengths of the home and away teams.
The strength of a team is defined as the average value of the ratings as
derived from each variant of our framework. Predictive performance for
the models is evaluated on the test set using proper scoring rules as
suggested by \citet{ArntzenHvattum21} and \citet{HvattumGelade21}. We
compute the Brier score \begin{align}
BS = \sum_{i = 1}^3 (d_i-p_i)^2,
\end{align} and the informational loss \begin{align}
IL = -\log_2\left(\sum_{i = 1}^3 p_id_i \right),
\end{align} where \(p_i\) is the model's probability of outcome
\(i = H,D,A\) and \(d_i = 1\), if match ended with outcome \(i\) and 0
otherwise. That is, for all matches in the test set, the Brier score
loss and the informational loss are obtained. The evaluation of a model
is done by considering the averages of these scores over the test
period. Furthermore, differences in the resulting averages can be
statistically tested using a two-sided paired \(t\)-test.

\par
\begin{table}[!htbp] \centering
  \caption{Average predictive losses from each strength model for the 100 matches in the test dataset under the different match prediction models and loss criteria.}
  \label{tab-val}
\begin{tabular}{@{\extracolsep{5pt}} lrrrr}
\\[-1.8ex]\hline
\hline \\[-1.8ex]
\hline \\[-1.8ex]
& \multicolumn{2}{c}{Ordinal Logistic} & \multicolumn{2}{c}{Bivariate Poisson} \\
Strength Model & BS & IL & BS & IL \\
\hline \\[-1.8ex]
Ridge Sum & $0.576$ & $1.398$ & $0.581$ & $1.406$ \\
Ridge Avg & $0.574$ & $1.392$ & $0.579$ & $\boldsymbol{1.403}$ \\
\hline \\[-1.8ex]
Group Lasso Sum & $\boldsymbol{0.566}$ & $\boldsymbol{1.381}$ & $\boldsymbol{0.578}$ & $1.404$ \\
Group Lasso Avg & $0.569$ & $1.387$ & $0.580$ & $1.410$ \\
\hline \\[-1.8ex]
Excl Lasso Sum & $0.577$ & $1.401$ & $0.589$ & $1.427$ \\
Excl Lasso Avg & $0.577$ & $1.403$ & $0.589$ & $1.427$ \\
\hline \\[-1.8ex]
Gen Lasso Sum & $0.611$ & $1.470$ & $0.618$ & $1.482$ \\
Gen Lasso Avg & $0.596$ & $1.443$ & $0.608$ & $1.465$ \\
\hline \\[-1.8ex]
PCV & $0.631$ & $1.514$ & $0.632$ & $1.514$ \\
ELO & $0.638$ & $1.525$ & $0.614$ & $1.466$ \\
Baseline & $0.636$ & $1.521$ & $0.637$ & $1.523$ \\
\hline \\[-1.8ex]
\end{tabular}
\end{table}

\begin{figure}
  \centering
  \includegraphics[width=\textwidth]{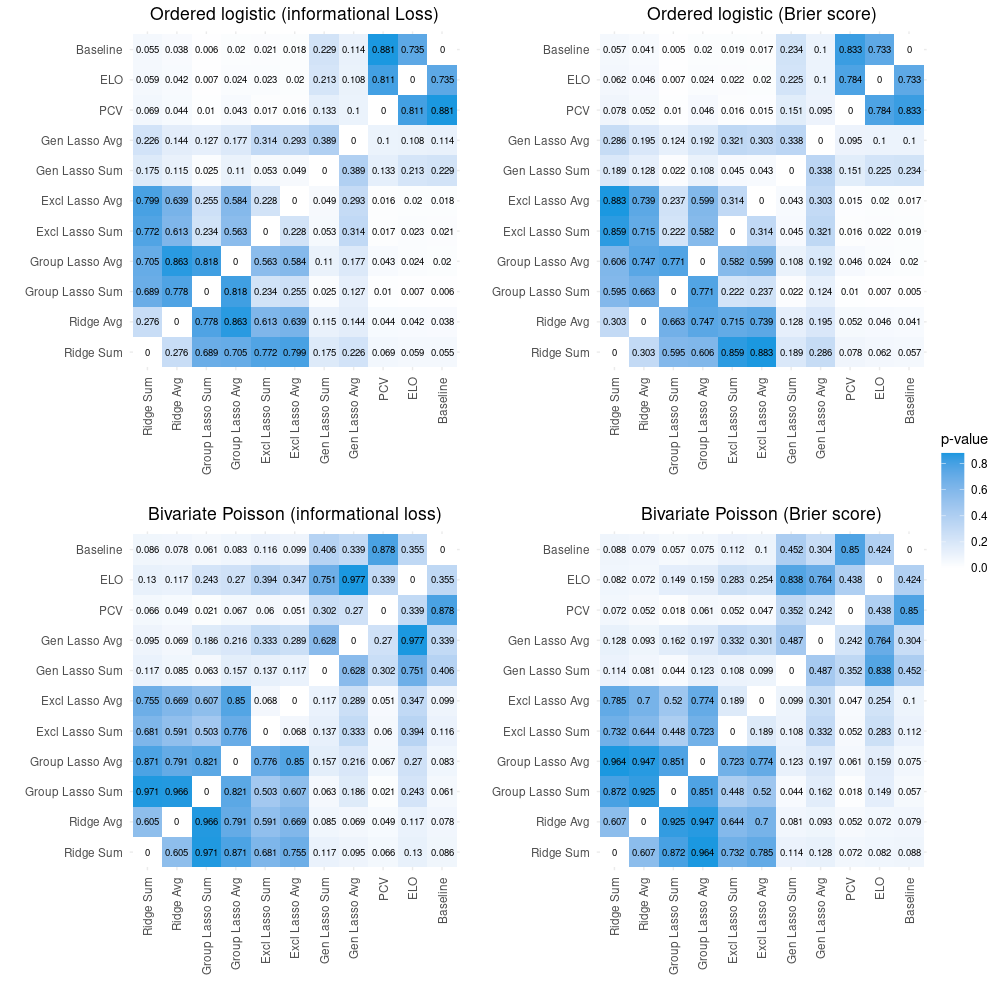}
  \caption{p-values of paired two-sided $t$-tests of the predictive losses of each model for 100 matches in test dataset under the different match prediction models and loss criteria.}
  \label{fig-p_vals_validity}
\end{figure}

Table \ref{tab-val} provides the results of the predictive performance
when using the respective strength estimate as a covariate. Furthermore,
three reference models are included. The model termed
\texttt{Baseline\textquotesingle{}\textquotesingle{}\ is\ a\ simple\ model\ without\ covariate,\ and\ the}ELO'\,'
model is derived by using ELO team ratings as taken from
\url{http://clubelo.com/}. The ``PCV'\,' model uses the \(PCV\) metric
described in \citet{Bajons23}. Details for the calculation of the
\(PCV\) value as well as a comparison of the metric with the ratings
described here are found in the supplementary material for this work.
The best-performing model in both prediction frameworks and under both
evaluation criteria seems to be the group lasso model. Although the
ridge regression model is able to keep up, it seems that there is value
in taking the positional information into account. The exclusive lasso
follows these two models closely, and the generalized lasso performs in
general worse than the aforementioned models. This result is not
unexpected as both these models are not necessarily optimal for deriving
a rating but rather have strengths in different use cases. However, all
of these models perform better in terms of validity than the three
reference models. Interestingly, the PCV and the ELO model both perform
only slightly better than the baseline model and are outperformed by the
regularized regression models. Figure \ref{fig-p_vals_validity} displays
a heatmap of the p-values of individual paired \(t\)-tests performed on
the predictive losses of the 100 matches in the test set. The results
from the \(t\)-test indicate that at least under the OLR model the best
performing models of the ridge, group lasso, and exclusive lasso penalty
significantly outperform the reference models (Baseline and ELO) at the
5 \% level. For the bivariate Poisson model, the differences to these
models are only significant at the 10 \% level, but the pattern is
similar. These results suggest that the regularized regression models
provide a rating that from an objective viewpoint captures the relevant
parts of the game well.

\section{Discussion}\label{sec-concl}

In this paper, we present a novel framework for evaluating the strength
of football players. We adapt the ideas of classical regularized
adjusted plus-minus models to possession sequences data and employ a
diverse set of penalization structures. The former adjustment allows for
evaluating on-ball as well as off-ball contributions of players -- an
often overlooked but important part of the game. The latter allows for
more flexibility and the incorporation of domain-specific
characteristics of the game by taking into account
e.g.\textasciitilde positions of players.

\par

In practice, each penalization structure has its advantages respectively
use cases. The general picture is, however, similar throughout all
presented models. The on-ball contributions are dominated by offensive
players (forwards and midfielders), while defensive players score better
on the off-ball contributions. From a domain-specific viewpoint, this
result is expected as the measurement for evaluating possession is
whether they result in a goal or not. It is reassuring that all the
methods act similarly in that sense and these results suggest that they
are meaningful and informative. Furthermore, we provide objective
justification for using our framework by conducting a
statistically-informed tailored-to-soccer validity analysis. The results
thereof suggest that using the group information of players provides
value and should be accounted for. Furthermore, the additional
information of on-ball and off-ball value can be leveraged in several
ways. For example, defensive players who perform well in on-ball
contributions may identify as offensively oriented defenders, which
might be desirable for specific styles of play. Similarly, offensive
players that demonstrate high off-ball value may have more impact on the
game than their active contributions (as measured through goals,
assists, etc.) are disclosing.

\par

The flexibility of our approach combined with the ease of adapting it to
existing procedures yields several possible future directions to boost
football player performance evaluation. For one, it might be interesting
to simply adapt existing regularized adjusted PM models to allow for the
presented regularization schemes. Also incorporating adjustments for
home advantage, manpower, league strength or relevancy of performance
(cf.~\citealp{Kharrat20,PH21}) is straightforward. However, this becomes
more relevant when more data is available. Another interesting direction
is to adapt the target variable and allow for some sort of hybrid
approach. One possibility is to follow \citet{Kharrat20} and use xG
values as target variable, however, any bottom-up rating that assigns a
value at the end of the possession may be used. One could even think of
ordinal-type target variables, however in this case the application of
the more elaborated penalization structures needs to be developed
further. Adapting the target may help discriminate between players, as
considering goals as outcome variables leads to a sparse and imbalanced
setup. Similarly, since analyzing possessions provides more detailed
insights into the game than analyzing segments as defined in PM models,
it is possible to adjust the outcome variable to allow for the valuation
of defensive actions such as the ability of a team to stop dangerous
possessions. Finally, future endeavors may explore and compare different
setups for incorporating player information. For example, it might be an
interesting avenue to account for the off-ball involvement by separating
offensive and defensive participation on the field (indirect
involvement). Figuratively speaking, a player like Lionel Messi, while
without a doubt positively influencing an offensive possession simply by
being a threat on the field, has arguably only little effect with regard
to defensive off-ball contribution.

\par

In summary, we believe that the presented framework allows for novel
perspectives and enhancements in the field of player evaluation and
scouting. The results are not only from a domain-specific but also from
an objective perspective useful, and thus the stakeholders involved in
the sports of football such as clubs, betting companies, media as well
as fans may benefit from the presented methods.

  \bibliography{References.bib}

\end{document}